\documentclass[useAMS,referee]{biom}
\usepackage{natbib}
\usepackage{graphicx}
\usepackage{amsmath}
\usepackage{amssymb}
\usepackage{dsfont}
\usepackage{multirow}
\usepackage{url}
\usepackage[figuresright]{rotating}
\usepackage{caption}
\usepackage{booktabs}
\usepackage{amsfonts}
\usepackage{xcolor}
\usepackage{mathtools}

\DeclareMathOperator*{\argmax}{argmax}
\DeclareMathOperator*{\argmin}{argmin}

\def\bSig\mathbf{\Sigma}

\title[Willard et. al. - Bayesian Optimization for Personalized Dose-Finding Trials]{Bayesian Optimization for Identification of Optimal Biological Dose Combinations in Personalized Dose-Finding Trials}

\author{James Willard$^{1,2,*}$\email{james.willard@mrc-bsu.cam.ac.uk}, 
Shirin Golchi$^{2}$, and Erica E.M. Moodie$^{2}$ \\
$^{1}$MRC Biostatistics Unit, University of Cambridge, Cambridge CB2 0SR, United Kingdom \\
$^{2}$Department of Epidemiology and Biostatistics, McGill University, Montreal H3A 1Y7, Canada}

\begin{document}





\pagerange{\pageref{firstpage}--\pageref{lastpage}} 
\volume{000}
\pubyear{2026}
\artmonth{August}



\label{firstpage}


\begin{abstract}
Early phase, personalized dose-finding trials for combination therapies seek to identify patient-specific optimal biological dose (OBD) combinations, which are defined as safe dose combinations that maximize therapeutic benefit for a specific covariate pattern. Given the small sample sizes which are typical of these trials, it is challenging for traditional parametric approaches to identify OBD combinations across multiple dosing agents and covariate patterns. To address these challenges, we propose a Bayesian optimization approach to dose-finding which incorporates efficacy and toxicity information into the sequential search strategy. Independent Gaussian processes are used to model the efficacy and toxicity surfaces, and an acquisition function is utilized to define the dose-finding strategy. Furthermore, we define an adaptive stopping rule using the posterior entropy for the location of the OBD. This work is motivated by a personalized dose-finding trial which considers a dual-agent therapy for obstructive sleep apnea (OSA), where OBD combinations are tailored to OSA severity. Via a simulation study, the approach is first investigated across varying degrees of response heterogeneity for both efficacy and toxicity, and then a collection of final designs for the OSA trial are compared. We demonstrate that the proposed approach toward personalized dose-finding yields good performance under the considered scenarios.
\end{abstract}

%



\maketitle


%

\section{Introduction}
\label{s:intro}

Early phase clinical trials assess the safety and efficacy of first-in-human doses of experimental therapies. Model-based Bayesian adaptive designs, which utilize posterior or posterior predictive distributions for sequential decision making, are commonly employed for these trials (e.g., \citealp{o1990continual}). Much of the historical work in early phase designs was motivated by the development of cytotoxic agents in oncology, where greater benefit was only possible with greater toxicity. These designs assumed binary responses and monotonic dose-response surfaces and sought to identify the largest safe dose, the so-called maximum tolerated dose (MTD; e.g., \citealp{o1990continual}). Recently, interest has grown in molecularly targeted compounds which may display benefit at dosages lower than the MTD \citep{le2009dose}. In these trials, traditional toxicity responses have been replaced by other drug-related biological effects, which are often measured on a continuous scale, e.g., plasma drug concentration and measures of target inhibition in tissues of interest, among others (e.g., \citealp{le2020adverse}). Despite this increased interest, dose-finding methods for continuous responses remains less developed than that for binary responses. Furthermore, the development of these novel therapies has shifted interest from identifying the MTD to identifying the dose with the best risk-benefit trade-off, the so-called optimal biological dose (OBD). Designs that target the OBD must incorporate both efficacy and toxicity information (e.g., \citealp{thall2004dose}) and may need to be flexible (e.g., \citealp{mozgunov2019flexible,mozgunov2019information}) to accommodate possible non-monotonic dose-response surfaces \citep{li2017toxicity}. 

All dose-finding designs mentioned above assume the same dose is optimal for every patient in the population. We refer to this ``one-size-fits-all'' approach as \textit{standard dose-finding}. With the continued identification and development of novel biomarkers, interest has grown in more personalized approaches to medicine. \textit{Personalized dose-finding} seeks to find optimal doses based on individual patient characteristics. Despite this growing interest, the literature for personalized dose-finding trials remains underdeveloped as the limited sample sizes in these trials make it challenging to extend many parametric dose-finding methods to the personalized setting, where potentially many dose-covariate interaction terms must be estimated. Personalized dose-finding trials for monotherapy have been explored previously (e.g., \citealp{babb2001patient, guo2022bayesian}). Our focus in this manuscript is on personalized dose-finding trials for \textit{combination therapies}, where the additional dosing agents increase the dimensionality and exacerbate the estimation challenges. \cite{willard2023bayesian} proposed a flexible personalized dose-finding design for combination therapies that assumed continuous responses. They showed how Bayesian optimization could be used to efficiently explore all dosing dimensions under no monotonicity assumptions. However, they assumed a minimal toxicity setting where every dose combination was assumed to be safe, and so toxicity was not considered during dose optimization. The present work focuses on generalizing the approach of \cite{willard2023bayesian} to the setting of higher-grade continuous toxicities. It does so by including a dose-escalation scheme and incorporating toxicity information into the search strategy employed by the Bayesian optimization based dose-finding method. Additionally, an adaptive stopping rule is proposed and estimation is performed in a fully Bayesian manner. We restrict our attention to dose-finding trials where the efficacy and toxicity responses can be evaluated contemporaneously, so that both sources of information may be used during dose optimization in an ``efficacy-integrated'' fashion \citep{yuan2024statistical}.

This work is motivated by a dose-finding design for a combination therapy which treats obstructive sleep apnea (OSA). OSA is a common sleep disorder that is estimated to affect more than 930 million adults worldwide \citep{schweitzer2023combination}. Continuous positive airway pressure (CPAP) is the standard treatment since it is very effective, but it is often poorly tolerated and results in low patient adherence \citep{schweitzer2023combination}. Currently, there is no approved pharmacotherapy for OSA, although a recent study has investigated the combination of an antimuscarinic agent (aroxybutynin) and a norepinephrine reuptake inhibitor (atomoxetine) as therapy \citep{schweitzer2023combination}.  While this study suggests that combination therapy is effective in reducing a key continuous endpoint, it revealed that potential response heterogeneity exists with respect to OSA severity and so a more targeted therapeutic approach would be beneficial. The current work aims to develop a design that can identify OBD combinations that are tailored to OSA severity.

The remainder of this manuscript is organized in the following manner. We first provide a brief introduction to Bayesian optimization and describe how to generalize the approach of \cite{willard2023bayesian} to the setting of higher-grade toxicities. This is followed by a simulation study where we investigate the performance of the proposed approach under a variety of scenarios. We then consider the previously described dose-finding design for a dual-agent therapy to treat OSA. We conclude with a discussion. 

\section{Bayesian Optimization for Dose-Finding}

\textbf{Personalized Dose-Finding} Consider a set of $J \in \mathbb{N}$ dosing agents. For each agent $j$ let $D^{\prime}_j=\{d^{\prime}_{j,1},d^{\prime}_{j,2},...,d^{\prime}_{j,c_j}\}$ be the set of $c_j$ discrete dose levels on their original scale. Now let $d_j\in(0,1)$ be a standardized dose level where the set of standardized dose levels for agent $j$ is $D_j=\{d_{j,1},d_{j,2},...,d_{j,c_j}\}$. Define $C=\prod_{j=1}^{J}|D_j|$ discrete dose combinations $\mathbf{d}=(d_1,..., d_J)\in \mathbb{D}$ where $\mathbb{D}$ is a discrete dose grid defined as the Cartesian product of the sets of the standardized levels of the agents, $\mathbb{D} = \bigtimes_{j=1}^{J}D_j$. Now consider a set of $P$ discrete covariates $\{Z_p\}_{p=1}^P$. For each covariate $p$, let $\mathcal{Z}_p$ be the set of its levels and define $K=\prod_{p=1}^{P}|\mathcal{Z}_p|$ strata as the Cartesian product of the sets of the covariate levels, $\mathbb{Z} = \bigtimes_{p=1}^{P}\mathcal{Z}_p$. Then, denote each stratum by $\mathbf{z}_k \in \mathbb{Z}$ for $k=1,...,K$. Personalized dose-finding seeks to identify optimal dose combinations for each of the $K$ strata. We refer to the continuous therapeutic function of interest (i.e., efficacy or utility) as the \textit{efficacy} function, denoted by $f(\mathbf{d},\mathbf{z}) \in \mathbb{R}$, and denote the continuous toxicity function by $g(\mathbf{d},\mathbf{z}) \in \mathbb{R}$.  We note that both functions are assumed to have been transformed such that smaller values denote being more desirable. Our goal is to find, for each stratum, the dose combination $\mathbf{d}$ which minimizes $f(\mathbf{d},\mathbf{z}_k)$, subject to a tolerable level of toxicity $g_k^{\dagger}$:
\begin{equation}
    \argmin_{\mathbf{d} \in \mathbb{D}} f(\mathbf{d},\mathbf{z}_k)\ \ \ \text{subject to}\ \ \ g(\mathbf{d},\mathbf{z}_k) \le g^{\dagger}_k \text{ for }\ \ k=1,...,K. 
\end{equation}
Unique toxicity thresholds $g^{\dagger}_k$ allow permissible toxicity to be defined with respect to specific covariate profiles. This is useful in settings where higher levels of toxicity may be permitted for more severe disease subtypes, which is the case for the OSA dose-finding trial that motivates this work. We solve the above minimization problem using Bayesian optimization, a derivative-free method which finds the global optima of expensive-to-evaluate objective functions \citep{garnett2023bayesian}. Bayesian optimization relies on stochastic surrogate models, commonly Gaussian Processes (GP), to estimate the efficacy and toxicity functions, and then employs an acquisition function to define a sequential search policy. It has recently been used for early phase dose-finding trials \citep{takahashi2021phase12, takahashi2021bayesian, willard2023bayesian}. Below, we generalize the dose-finding approach proposed in \cite{willard2023bayesian} to higher-grade toxicity settings by using Bayesian optimization under toxicity constraints.


Consider a dose-finding trial divided into discrete time points denoted by $t=0,1,...,T$. Dose finding proceeds sequentially, where patient responses through time $t$ are used to inform the assignment of patients at time $t+1$. For safety reasons, dose-finding under higher-grade toxicity starts at the lowest dose and successively evaluates larger doses according to a dose escalation scheme that is described below. For general time $t$, let $r_{k,t}$ be the number of patient responses in stratum $k$ observed at dose $\mathbf{d}$, and $n_t=\sum_{k=1}^{K} r_{k,t}$ be the total number of patient responses. These responses yield noisy observations of both the efficacy function, $y_{f} = f(\mathbf{d},\mathbf{z}) + \epsilon_{f}$, and the toxicity function,  $y_{g} = g(\mathbf{d},\mathbf{z}) + \epsilon_{g}$, where $\epsilon_{f} \sim N(0,\sigma^2_{y_f})$, $\epsilon_{g} \sim N(0,\sigma^2_{y_g})$. We denote the vector of responses for the patient $i$ by $\mathbf{y}_i=(y_{f}, y_{g})_i^T$, the vector of efficacy responses for all patients by $\mathbf{y}_{f,t}=(y_{f,1},..., y_{f,n_t})^T$, and the vector of toxicity responses for all patients by $\mathbf{y}_{g,t}=(y_{g,1},..., y_{g,n_t})^T$. It is common to normalize the responses and we assume this has been done below. In many contexts, it is reasonable to assume conditional independence between the efficacy and toxicity functions given dose and covariates, and we do so throughout. As our motivating trial suggests potentially non-monotonic dose-response surfaces, we utilize independent, unconstrained GP models for the efficacy and toxicity functions, the details of which are described below for the efficacy function only since those for toxicity are similar. Thus, a GP prior is assumed for the efficacy function, 
\begin{equation}
    f(\mathbf{d}, \mathbf{z}) \sim GP\left(m_f(\mathbf{d},\mathbf{z}), \nu^2_f\mathcal{K}_f((\mathbf{d},\mathbf{z)}, (\mathbf{d}^{\prime}, \mathbf{z}^{\prime}))\right) 
\end{equation}
where $m_f(\mathbf{d},\mathbf{z})$ is the mean function, and $\mathcal{K}_f((\mathbf{d},\mathbf{z}), (\mathbf{d}^{\prime},\mathbf{z}^{\prime}))$ is a correlation function (kernel) multiplied by scale parameter $\nu^2_f$. The scale parameter determines the variability of the efficacy function throughout the dose combination space. We utilize GP models with mean functions $m_f(\mathbf{d},\mathbf{z}) = \beta_{f}(\mathbf{d},\mathbf{z})$, which provide prior estimates of the dose-response surfaces and help the design better explore the dose combination space. These are similar to the ``dose-skeletons'' (see e.g., \citealp{o1990continual}) that have traditionally been used in early phase dose-finding designs. Note that information about the efficacy function (e.g., pharmacokinetic/pharmacodynamic models) or toxicity function (e.g., quantitative systems toxicology models) could also be incorporated into the mean functions. Throughout this work, we assume stationary dose-response surfaces, where the degree of correlation between two dose combinations depends only on their distance from one another in the input space. We employ a stationary anisotropic squared exponential kernel,
\begin{equation}\label{cov_squared_exp_kernel}
     \mathcal{K}_f((\mathbf{d},\mathbf{z}),(\mathbf{d}^{\prime},\mathbf{z}^{\prime}))=\exp\left\{-\left(\sum_{j=1}^{J}\frac{(d_j - d_j^{\prime})^2}{2l_{f,d_j}^2} + \sum_{p=1}^{P}\frac{(z_p - z_p^{\prime})^2}{2l_{f,z_p}^2}\right)\right\},
\end{equation}
which is parameterized by characteristic length-scales $\{l_{f,d_1},...,l_{f,z_P}\}$. The length-scales control how quickly the correlation between two dose combinations decays with respect to each dosing agent and across covariate patterns.

Specifying this GP prior induces a multivariate normal distribution on the efficacy function observations, 
$\mathbf{y}_{f} \sim N(\boldsymbol{\beta}_f, \mathbf{K}_f)$, where $\mathbf{K}_f(i,j)=\nu^2_f\mathcal{K}_f((\mathbf{d}_i,\mathbf{z}_i),(\mathbf{d}_j,\mathbf{z}_j)) + \tau_f^2\mathds{1}_{i=j}$ with noise parameter $\tau_f^2$, and where $\boldsymbol{\beta}_f=(\beta_f(\mathbf{d}_i,\mathbf{z}_i),...,\beta_f(\mathbf{d}_n,\mathbf{z}_n))$. After data $\mathcal{D}_t = \{(\mathbf{d}_i,\mathbf{z}_i,\mathbf{y}_i)\}_{i=1}^{n_t}$ are observed, the joint posterior distribution of the kernel hyperparameters $\boldsymbol{\theta}_f = \{\nu_f, \tau_f, l_{f,d_1},..., l_{f,z_P}\}$ is estimated under assumed prior distributions. See Section S2 of the Supplementary Materials for a suggested approach in specifying the priors on $\boldsymbol{\beta}_f$ and $\boldsymbol{\theta}_f$.

This yields the posterior distribution of the efficacy function for a new dose combination $\widetilde{\mathbf{d}}$ in stratum $k$, denoted by $\widetilde{\mathbf{d}}_k = (\widetilde{\mathbf{d}}, \widetilde{\mathbf{z}}_k)$, as $p(f \mid \mathcal{D}_t, \widetilde{\mathbf{d}}_k) = N(\mu_f(\widetilde{\mathbf{d}}_k), \sigma_f^2(\widetilde{\mathbf{d}}_k))$ \citep{garnett2023bayesian}, where $S$ posterior samples of $\mu_{f}(\widetilde{\mathbf{d}}_k)$ and $\sigma_f^{2}(\widetilde{\mathbf{d}}_k)$ are obtained as
\begin{equation}\label{posterior}
\begin{aligned}
    \mu_{f}^{(s)}(\widetilde{\mathbf{d}}_k) &= \beta_f(\widetilde{\mathbf{d}}_k) + \mathbf{k}_f^{(s)}(\widetilde{\mathbf{d}}_k)^{T}\mathbf{K}_f^{-1(s)}(\mathbf{y}_f - \beta_f(\widetilde{\mathbf{d}}_k))\\
    \sigma_f^{(s)}(\widetilde{\mathbf{d}}_k) &= \left[\nu^2_f\mathcal{K}_f^{(s)}(\widetilde{\mathbf{d}}_k,\widetilde{\mathbf{d}}_k) - \mathbf{k}_f^{(s)}(\widetilde{\mathbf{d}}_k)^{T}\mathbf{K}_f^{-1(s)}\mathbf{k}_f^{(s)}(\widetilde{\mathbf{d}}_k) \right]^{\frac{1}{2}}
\end{aligned}
\end{equation}
where $\mathbf{K}_f^{(s)}$ is $n \times n$ and $\mathbf{k}_f^{(s)}(\widetilde{\mathbf{d}}_k)=[\nu^2_f\mathcal{K}_f^{(s)}((\mathbf{d}_1, \mathbf{z}_1),\widetilde{\mathbf{d}}_k),..., \nu^2_f\mathcal{K}_f^{(s)}((\mathbf{d}_n, \mathbf{z}_n),\widetilde{\mathbf{d}}_k)]^{T}$ is $n \times 1$, each of which is calculated from the $s^{th}$ sample for $s=1,...,S$ from the joint posterior distribution of $\boldsymbol{\theta}_f$. The posterior distribution of the toxicity function at $\widetilde{\mathbf{d}}_k$ is obtained similarly and has the same functional form as above but where $f$ is replaced by $g$ for all observations and kernel hyperparameters. To ease notation, we will refer to the collection of joint posterior samples as $\boldsymbol{\theta}=\{\boldsymbol{\theta}_f,\boldsymbol{\theta}_g\}$.

The next dose combination for evaluation within stratum $k$, denoted by $\mathbf{d}_k^{(t+1)}$, is selected from an admissible set of doses at time $t$, denoted by $\mathcal{A}_{k,t} \subseteq \mathbb{D}$. At the start of the trial at time $t=0$, the smallest dose, defined as the $J$-dimensional vector $\mathbf{d}_k=\mathbf{0}$, is evaluated first. Thereafter, a dose-escalation scheme sequentially evaluates increasingly larger doses that are potentially more toxic. In a combination therapy setting, many doses may yield comparable toxicity and so there is not a natural ordering of doses. Thus, dose-escalation proceeds by permitting the region of admissible doses $\mathcal{A}_{k,t}$ to sequentially expand at a rate of $\rho_k$ in each coordinate direction according to the following:
\begin{equation}
    \mathcal{A}_{k,t}=\{\mathbf{d}_k: \mathbf{d}_k \in [0,\rho_k\times (t+1)]^J \cap \mathbb{D}, \rho_k \in (0,1)\}.
\end{equation}
For $J=2$, this would equate to permitting dose-finding to proceed ``along the diagonal'', which has been shown to lead to more efficient identification of the MTD in some cases \citep{sweeting2012escalation}. We note that $\rho_k$ is a tuning parameter that may impact performance and so can be calibrated through simulation.

The next dose combination $\mathbf{d}_k^{(t+1)}$ is then selected as the dose $\widetilde{\mathbf{d}}_k \in \mathcal{A}_{k,t}$ that maximizes an acquisition function, denoted by $\alpha(\widetilde{\mathbf{d}}_k)$: $\mathbf{d}_k^{(t+1)} = \argmax_{\widetilde{\mathbf{d}}_k \in \mathcal{A}_{k,t}} \alpha(\widetilde{\mathbf{d}}_k)$. One acquisition function which can be used in the presence of toxicity constraints is the Expected Constrained Improvement (cEI; \citealp{gardner2014bayesian}) defined as
\begin{equation}
    \alpha_{cEI,\boldsymbol{\theta},f_k^*}(\widetilde{\mathbf{d}}_k) = \mathbb{E}_{f,g}\left[\max(0,f_k^*-f(\widetilde{\mathbf{d}}_k))\mathds{1}\{g(\widetilde{\mathbf{d}}_k) \le g_k^{\dagger}\} \mid \mathcal{D}_t,\widetilde{\mathbf{d}}_k,\boldsymbol{\theta},f_k^*\right],
\end{equation}  
where we have made explicit the conditioning on a specific value of $\boldsymbol{\theta}$ and $f_k^*$, the current optimum of the efficacy function in stratum $k$. Under a noisy setting, $f_k^*$ is not observed and must be estimated. Given a single sample from the joint posterior of $\boldsymbol{\theta}$, we obtain a single posterior sample of $f_k^{*}$ as $f_k^{*(s)}=\min_{\widetilde{\mathbf{d}}_k \in \mathcal{A}_{k,t}}\{f^{(s)}(\widetilde{\mathbf{d}}_k)\ \text{ subject to }\ g^{(s)}(\widetilde{\mathbf{d}}_k) \le g_k^{\dagger}\}$. The expectation above is available in closed form under the assumption of conditional independence between the efficacy and toxicity functions given dose and covariates \citep{gardner2014bayesian}, which yields the following for the $s^{th}$ posterior sample of $\boldsymbol{\theta}$:
\begin{equation}
\begin{aligned}
     \alpha_{cEI}^{(s)}(\widetilde{\mathbf{d}}_k) = &\left[(f_k^{*(s)} - \mu_f^{(s)}(\widetilde{\mathbf{d}}_k))\Phi\left(\frac{f_k^{*(s)}-\mu_f^{(s)}(\widetilde{\mathbf{d}}_k)}{\sigma_f^{(s)}({\widetilde{\mathbf{d}}_k)}}\right) + \sigma_f^{(s)}(\widetilde{\mathbf{d}}_k)\phi\left(\frac{f_k^{*(s)}-\mu_f^{(s)}(\widetilde{\mathbf{d}}_k)}{\sigma_f^{(s)}({\widetilde{\mathbf{d}}_k)}}\right)\right] \times \\ &\Phi\left(\frac{g_k^{\dagger} - \mu_g^{(s)}(\widetilde{\mathbf{d}}_k)}{\sigma_g^{(s)}(\widetilde{\mathbf{d}}_k)}\right). 
\end{aligned}
\end{equation}
Above, $g_k^{\dagger}$ denotes the stratum-specific toxicity constraint, $\Phi(\cdot)$ and $\phi(\cdot)$ denote the standard normal cumulative distribution function and probability density function, respectively, and $\mu^{(s)}(\widetilde{\mathbf{d}}_k)$ and $\sigma^{(s)}(\widetilde{\mathbf{d}}_k)$ are given above in (\ref{posterior}). The expression inside the brackets serves to balance the trade-off between exploring regions of the dose combination space where the efficacy function is imprecisely estimated, and exploiting regions which have desirable values of the efficacy function. This trade-off is weighted by the posterior probability of satisfying the toxicity constraint, giving higher weight to points which are more likely to be safe. We use $S$ posterior samples to approximate an integrated form of cEI, denoted by $\alpha_{cEI}(\widetilde{\mathbf{d}}_k)$, that fully accounts for the uncertainty in both $f_k^{*}$ and $\boldsymbol{\theta}$:
\begin{equation}
    \alpha_{cEI}(\widetilde{\mathbf{d}}_k) = \int \alpha_{cEI,\boldsymbol{\theta},f_k^*}(\widetilde{\mathbf{d}}_k)p(f_k^*,\boldsymbol{\theta} \mid \mathcal{D}_t)df_k^*d\boldsymbol{\theta} \approx \frac{1}{S}\sum_{s=1}^{S} \alpha_{cEI}^{(s)}(\widetilde{\mathbf{d}}_k).
\end{equation}
After observing responses for $r_{k,t}$ new patients at $\mathbf{d}_k^{(t+1)}$, the data are updated and the GP models are refit to obtain new posterior distributions for $f$ and $g$. Then $S$ samples $f^{(s)}(\widetilde{\mathbf{d}}_k)$ and $g^{(s)}(\widetilde{\mathbf{d}}_k)$ from these posteriors are obtained to yield $S$ samples from the posterior of the optimal dose combination $p(\mathbf{d}_{opt,k} \mid \mathcal{D}_t)$ as:
\begin{equation}
    \mathbf{d}^{(s)}_{opt,k} =\argmin_{\widetilde{\mathbf{d}}_k \in \mathcal{A}_{k,t}}\{f^{(s)}(\widetilde{\mathbf{d}}_k)\ \text{ subject to }\ g^{(s)}(\widetilde{\mathbf{d}}_k) \le g_k^{\dagger}\}.
\end{equation}
Dose-finding within each stratum continues until the sample size limit is reached or a stratum-specific early stopping rule is satisfied. If we let $p_k(\mathbf{d}) = p(\mathbf{d}_{opt,k} \mid \mathcal{D}_t)$, then we permit early stopping for efficacy in stratum $k$ when the entropy of $p_k(\mathbf{d})$ is less than $\delta_k$ for at least $(J+1)$ consecutive time points, i.e., when $H_k=\Sigma_{\mathbf{d}\in\mathbb{D}}p_k(\mathbf{d})\log p_k(\mathbf{d})<\delta_k$ for at least $(J+1)$ consecutive time points, noting that $p_k(\mathbf{d})\log p_k(\mathbf{d})$ is defined to be 0 if $p_k(\mathbf{d})=0$. Intuitively, stopping occurs when enough posterior mass has concentrated on a promising region of the dose combination space and this must occur for several time points so we are confident it has not resulted from random variability in the data. The value of $\delta_k$ is calibrated through sensitivity analysis. For convenience, we refer to this as early stopping for ``efficacy'', though note this need not imply a significant therapeutic response. Similarly, stratum-specific early stopping for toxicity may occur if no dose satisfies $P(g(\mathbf{d}_k) \le g_k^{\dagger} \mid \mathcal{D}_t) > \gamma_k$ (i.e., no safe doses) for at least $(J+1)$ consecutive time points, where $\gamma_k$ is a tuning parameter. At the end of dose optimization, the patient-specific optimal dose combinations are selected as the posterior modes of $p(\mathbf{d}_{opt,k} \mid \mathcal{D}_t)$. The procedure described above permits \textit{personalized dose-finding}. We note this method accommodates \textit{standard dose-finding} by ignoring covariate information (i.e., $\{Z_p\}_{p=1}^P= \emptyset$), though we do not consider standard dose-finding further in this work.

\section{Simulation Study}

\begin{figure}
    \centering
    \includegraphics[width = \linewidth]{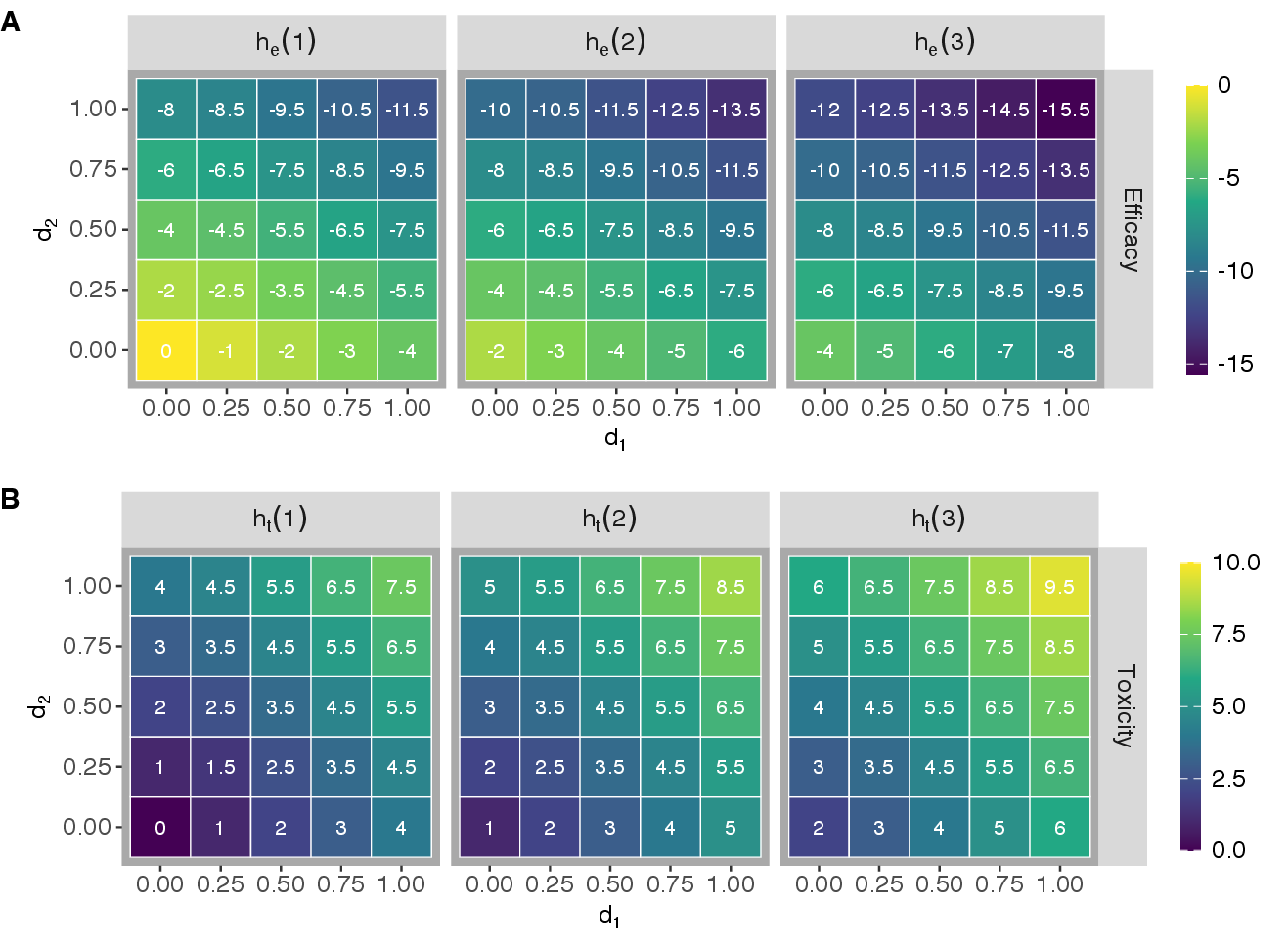}
    \caption{A) Efficacy and B) toxicity settings used to define $f(\mathbf{d},z_1)$ and $g(\mathbf{d},z_1)$ for each scenario in the simulation study described in Table \ref{tbl:sim_study_scenarios}. Each efficacy setting is indicated by $h_e(i)$ and each toxicity setting is indicated by $h_t(i)$ for $i=1,2,3$. The settings become increasingly efficacious/toxic as $i$ increases. The yellow-green end of the color spectrum denotes worse efficacy and toxicity and the dark blue end denotes better efficacy and toxicity.}
    \label{fig:sim_study_eff_tox_settings}
\end{figure}

\begin{figure}
    \centering
    \includegraphics[width = \linewidth]{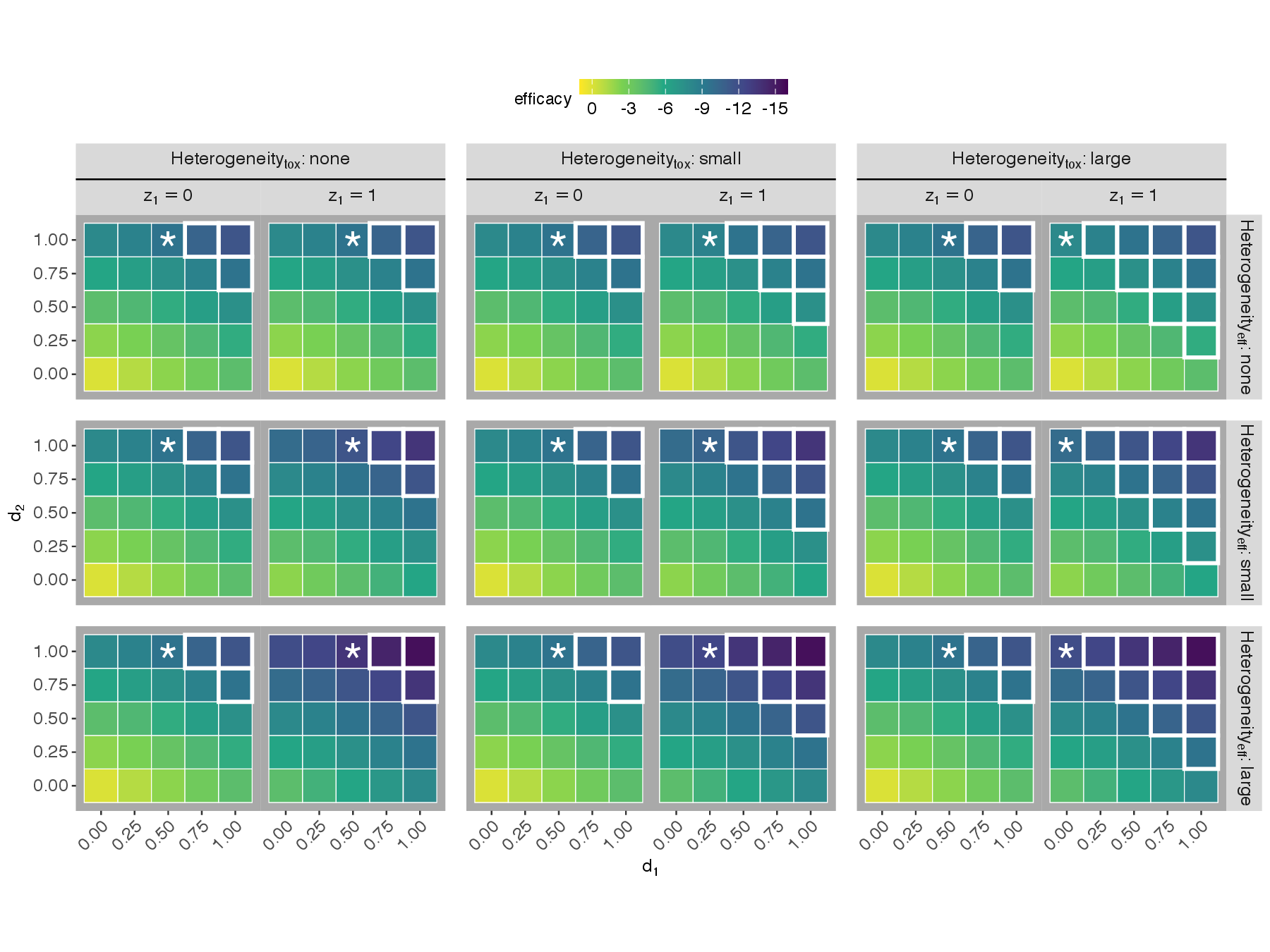}
    \caption{Simulation study scenarios as defined in Table \ref{tbl:sim_study_scenarios}. Nine scenarios are formed by combining three levels of heterogeneity in efficacy (none, small, large) denoted by $\text{Heterogeneity}_{eff}$ in the right facet labels, with three levels of heterogeneity in toxicity (none, small, large) denoted by $\text{Heterogeneity}_{tox}$ in the top facet labels. White stars denote $\mathbf{d}_{opt,k}$ and white borders around cells denote toxic doses, which are doses where $g(\mathbf{d},z_1) > 6$. The yellow-green end of the color spectrum denotes worse efficacy and the dark blue end denotes better efficacy.}
    \label{fig:sim_study_all_scenarios}
\end{figure}

In this section, we perform a simulation study to investigate the performance of the approach for a dual-agent (i.e., $J=2$) therapy under a variety of scenarios which consider a single binary covariate $z_1$. We utilize a set of 25 dose combinations $\mathbf{d}=(d_1,d_2)$, defined over the Cartesian product of $d_1,d_2 \in \{0,0.25,0.5,0.75,1\}$. We set the maximum sample size to 120 participants, $r_{k,t}=2$ participants for all $t$, and $\rho_k=0.5$. We consider three settings for each of efficacy and toxicity (Figure \ref{fig:sim_study_eff_tox_settings}) which are combined to yield nine simulation scenarios that vary in the degree of response heterogeneity across $z_1$ (Table \ref{tbl:sim_study_scenarios} and visualized in Figure \ref{fig:sim_study_all_scenarios}). We use values of $z_1$ to index the true optimal dose combinations, $\mathbf{d}_{opt,k}$, and the values of the efficacy and toxicity functions at the $\mathbf{d}_{opt,k}$, denoted by $f_{opt,k}$ and $g_{opt,k}$, respectively. At the end of dose-finding, the recommended dose is $\widehat{\mathbf{d}}_{opt,k}=\argmax_{\mathbf{d}_k}p(\mathbf{d}_{opt,k} \mid \mathcal{D}_t)$.

Within the simulation scenarios, we have three degrees of response heterogeneity for toxicity (none, small, large), and three degrees of response heterogeneity for efficacy (none, small, large), which are denoted by $\text{Heterogeneity}_{tox}$: none/small/large and $\text{Heterogeneity}_{eff}$: none/small/large, respectively. Figure \ref{fig:sim_study_all_scenarios} displays the nine simulation scenarios which are separated by white space. Within each scenario, thick white borders around cells are used to denote toxic doses, which is any dose $\mathbf{d}_k$ whose true toxicity is greater than 6 (i.e., $g_k^{\dagger}=6$). We do not consider a scenario where all dose combinations are toxic, since the dose ranges investigated in combination therapies are expected to be well targeted due to the previous early phase studies of each individual agent \citep{sweeting2012escalation}. Within a specific efficacy scenario (right facet label), the location of $\mathbf{d}_{opt,k}$ (white star) changes across $z_1$ as we move from no to large response heterogeneity in toxicity (left to right in the top facet label). Within a specific toxicity scenario (top facet label), the location of $\mathbf{d}_{opt,k}$ remains the same but the efficacy surface changes as we move from no to large response heterogeneity in efficacy (top to bottom in right facet label). The scenarios do not permit early stopping for efficacy, which is investigated in the next section. Furthermore, we do not permit early stopping for toxicity as we expect the dose combinations to be well targeted and are willing to assume there is at least one safe dose.

\begin{table}
    \caption{Simulation scenarios considered.  The data generating mechanism for each scenario is $y_f = f(\mathbf{d},z_1) + \epsilon_f$ and $y_g = g(\mathbf{d},z_1) + \epsilon_g$ where $\epsilon_f \sim N(0, \sigma^2_{y_f})$ and $\epsilon_g \sim N(0, \sigma^2_{y_g})$ and the values for $h_e$ and $h_t$ come from Figure \ref{fig:sim_study_eff_tox_settings}. The table columns contain the location of the optimal dose combination ($\mathbf{d}_{opt}$), the value of the efficacy and toxicity functions at $\mathbf{d}_{opt}$ ($f_{opt}/g_{opt}$), and the standardized effect sizes ($ses_f/ses_g$).}
    \centering
    \resizebox{\linewidth}{!}{
   \begin{tabular}[width=\linewidth]{clllllllllc}
    \toprule
    & $f(\mathbf{d},z_1)$ & $g(\mathbf{d},z_1)$ & $\sigma_{y_f}$ & $\sigma_{y_g}$ & $z_1$ & $\mathbf{d}_{opt}$ & $f_{opt}$ & $g_{opt}$ & $ses_f$ & $ses_g$\\
    \midrule 
    \raisebox{0.5em}{\multirow{30}{*}{\rotatebox[origin=r]{90}{{Simulation Study}}}}

    & \multirow{3}{*}{\parbox{0.3\linewidth}{$\textbf{Heterogeneity}_{\textbf{eff}}\textbf{: none} \\ \mathds{1}\{z_1=0\} \times h_e(1) + \\ \mathds{1}\{z_1=1\} \times h_e(1) $}} &
    \multirow{3}{*}{\parbox{0.3\linewidth}{$\textbf{Heterogeneity}_{\textbf{tox}}\textbf{: none} \\ \mathds{1}\{z_1=0\} \times h_t(1)  + \\  \mathds{1}\{z_1=1\} \times h_t(1) $}}
     & 9.5 & 5.5 & 0 & $(0.5, 1.0)$ & -9.5 & 5.5 & 1 & 1\\
    & & & & & 1 & $(0.5, 1.0)$ &  -9.5 & 5.5 & 1 & 1\\ \\
    \cmidrule{2-11}

    & \multirow{3}{*}{\parbox{0.25\linewidth}{$\textbf{Heterogeneity}_{\textbf{eff}}\textbf{: none} \\ \mathds{1}\{z_1=0\} \times h_e(1) + \\ \mathds{1}\{z_1=1\} \times h_e(1) $}} &
    \multirow{3}{*}{\parbox{0.25\linewidth}{$\textbf{Heterogeneity}_{\textbf{tox}}\textbf{: small} \\ \mathds{1}\{z_1=0\} \times h_t(1)  + \\  \mathds{1}\{z_1=1\} \times h_t(2) $}}
     & 9.5 & 5.5 & 0 & $(0.5, 1.0)$ & -9.5 & 5.5 & 1 & 1\\
    & & & & & 1 & $(0.25, 1.0)$ &  -8.5 & 5.5 & 0.89 & 1\\ \\
    \cmidrule{2-11}

    & \multirow{3}{*}{\parbox{0.25\linewidth}{$\textbf{Heterogeneity}_{\textbf{eff}}\textbf{: none} \\ \mathds{1}\{z_1=0\} \times h_e(1) + \\ \mathds{1}\{z_1=1\} \times h_e(1) $}} &
    \multirow{3}{*}{\parbox{0.25\linewidth}{$\textbf{Heterogeneity}_{\textbf{tox}}\textbf{: large} \\ \mathds{1}\{z_1=0\} \times h_t(1)  + \\  \mathds{1}\{z_1=1\} \times h_t(3) $}}
     & 9.5 & 5.5 & 0 & $(0.5, 1.0)$ & -9.5 & 5.5 & 1 & 1\\
    & & & & & 1 & $(0.0, 1.0)$ &  -8.0 & 6.0 & 0.84 & 1.09\\ \\
    \cmidrule{2-11}

    & \multirow{3}{*}{\parbox{0.25\linewidth}{$\textbf{Heterogeneity}_{\textbf{eff}}\textbf{: small} \\ \mathds{1}\{z_1=0\} \times h_e(1) + \\ \mathds{1}\{z_1=1\} \times h_e(2) $}} &
    \multirow{3}{*}{\parbox{0.25\linewidth}{$\textbf{Heterogeneity}_{\textbf{tox}}\textbf{: none} \\ \mathds{1}\{z_1=0\} \times h_t(1)  + \\  \mathds{1}\{z_1=1\} \times h_t(1) $}}
     & 9.5 & 5.5 & 0 & $(0.5, 1.0)$ & -9.5 & 5.5 & 1 & 1\\
    & & & & & 1 & $(0.5, 1.0)$ &  -11.5 & 5.5 & 1.21 & 1\\ \\
    \cmidrule{2-11}

    & \multirow{3}{*}{\parbox{0.25\linewidth}{$\textbf{Heterogeneity}_{\textbf{eff}}\textbf{: small} \\ \mathds{1}\{z_1=0\} \times h_e(1) + \\ \mathds{1}\{z_1=1\} \times h_e(2) $}} &
    \multirow{3}{*}{\parbox{0.25\linewidth}{$\textbf{Heterogeneity}_{\textbf{tox}}\textbf{: small} \\ \mathds{1}\{z_1=0\} \times h_t(1)  + \\  \mathds{1}\{z_1=1\} \times h_t(2) $}}
     & 9.5 & 5.5 & 0 & $(0.5, 1.0)$ & -9.5 & 5.5 & 1 & 1\\
    & & & & & 1 & $(0.25, 1.0)$ &  -10.5 & 5.5 & 1.11 & 1\\ \\
    \cmidrule{2-11}

    & \multirow{3}{*}{\parbox{0.25\linewidth}{$\textbf{Heterogeneity}_{\textbf{eff}}\textbf{: small} \\ \mathds{1}\{z_1=0\} \times h_e(1) + \\ \mathds{1}\{z_1=1\} \times h_e(2) $}} &
    \multirow{3}{*}{\parbox{0.25\linewidth}{$\textbf{Heterogeneity}_{\textbf{tox}}\textbf{: large} \\ \mathds{1}\{z_1=0\} \times h_t(1)  + \\  \mathds{1}\{z_1=1\} \times h_t(3) $}}
     & 9.5 & 5.5 & 0 & $(0.5, 1.0)$ & -9.5 & 5.5 & 1 & 1\\
    & & & & & 1 & $(0.0, 1.0)$ &  -10.0 & 6.0 & 1.05 & 1.09\\ \\
    \cmidrule{2-11}

    & \multirow{3}{*}{\parbox{0.25\linewidth}{$\textbf{Heterogeneity}_{\textbf{eff}}\textbf{: large} \\ \mathds{1}\{z_1=0\} \times h_e(1) + \\ \mathds{1}\{z_1=1\} \times h_e(3) $}} &
    \multirow{3}{*}{\parbox{0.25\linewidth}{$\textbf{Heterogeneity}_{\textbf{tox}}\textbf{: none} \\ \mathds{1}\{z_1=0\} \times h_t(1)  + \\  \mathds{1}\{z_1=1\} \times h_t(1) $}}
     & 9.5 & 5.5 & 0 & $(0.5, 1.0)$ & -9.5 & 5.5 & 1 & 1\\
    & & & & & 1 & $(0.5, 1.0)$ &  -13.5 & 5.5 & 1.42 & 1\\ \\
    \cmidrule{2-11}

    & \multirow{3}{*}{\parbox{0.25\linewidth}{$\textbf{Heterogeneity}_{\textbf{eff}}\textbf{: large} \\ \mathds{1}\{z_1=0\} \times h_e(1) + \\ \mathds{1}\{z_1=1\} \times h_e(3) $}} &
    \multirow{3}{*}{\parbox{0.25\linewidth}{$\textbf{Heterogeneity}_{\textbf{tox}}\textbf{: small} \\ \mathds{1}\{z_1=0\} \times h_t(1)  + \\  \mathds{1}\{z_1=1\} \times h_t(2) $}}
     & 9.5 & 5.5 & 0 & $(0.5, 1.0)$ & -9.5 & 5.5 & 1 & 1\\
    & & & & & 1 & $(0.25, 1.0)$ &  -12.5 & 5.5 & 1.32 & 1\\ \\
    \cmidrule{2-11}

    & \multirow{3}{*}{\parbox{0.25\linewidth}{$\textbf{Heterogeneity}_{\textbf{eff}}\textbf{: large} \\ \mathds{1}\{z_1=0\} \times h_e(1) + \\ \mathds{1}\{z_1=1\} \times h_e(3) $}} &
    \multirow{3}{*}{\parbox{0.25\linewidth}{$\textbf{Heterogeneity}_{\textbf{tox}}\textbf{: large} \\ \mathds{1}\{z_1=0\} \times h_t(1)  + \\  \mathds{1}\{z_1=1\} \times h_t(3) $}}
     & 9.5 & 5.5 & 0 & $(0.5, 1.0)$ & -9.5 & 5.5 & 1 & 1\\
    & & & & & 1 & $(0.0,1.0)$ &  -12.0 & 6.0 & 1.26 & 1.09\\ \\


    \midrule 
    
    \raisebox{-0.5em}{\multirow{2}{*}{\rotatebox[origin=c]{90}{{OSA}}}}
    & \multirow{3}{*}{\parbox{0.25\linewidth}{$\phi_{z_1,0} + \phi_{z_1,1} d_1 + \phi_{z_1,2} d_2 + \\ \phi_{z_1,3} d_1 d_2 + \phi_{z_1,4} d_1^2 + \\ \phi_{z_1,5} d_2^2 + \phi_{z_1,6} d_1^2 d_2^2$}} &
    \multirow{3}{*}{\parbox{0.2\linewidth}{$\psi_0 + \psi_1d_1 + \psi_2d_2 + \\ \psi_3 d_1 d_2 + \psi_4 d_1^2 + \\ \psi_5 d_2^2 + \psi_6 d_1^2 d_2^2$}} & 7.68 & 1.29 & 0 & $(2.5, 75)$ & -7.68 & 1.29 & 1 & 1\\
    &  &  &  & & 1 & $(5, 75)$ & -13.20 & 1.63 & 1.72 & 1.26 \\
    \nonumber \\  
    \bottomrule
    
    
    \end{tabular}
    }
    \label{tbl:sim_study_scenarios}
\end{table}

\begin{figure}
    \centering
    \includegraphics[width=\linewidth]{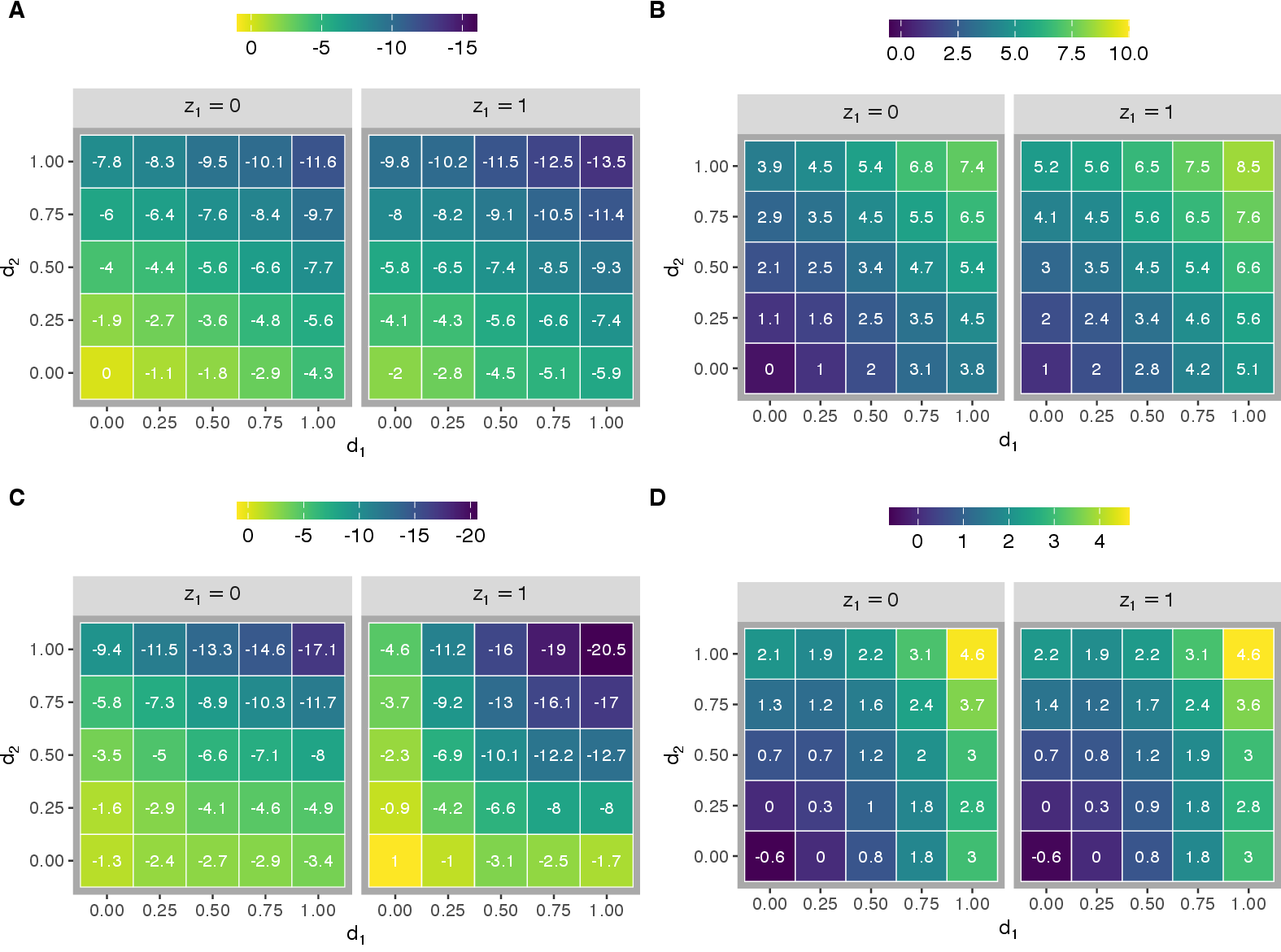}
    \caption{A) Prior mean $\boldsymbol{\beta}_f$ used in the simulation study, B) prior mean $\boldsymbol{\beta}_g$ used in the simulation study, C) prior mean $\boldsymbol{\beta}_f$ used in the OSA trial, D) prior mean $\boldsymbol{\beta}_g$ used in used in the OSA trial. The yellow-green end of the color spectrum denotes worse efficacy and toxicity and the dark blue end denotes better efficacy and toxicity.}
    \label{fig:prior_mean}
\end{figure}

\begin{figure}
    \centering
    \includegraphics[width = \linewidth]{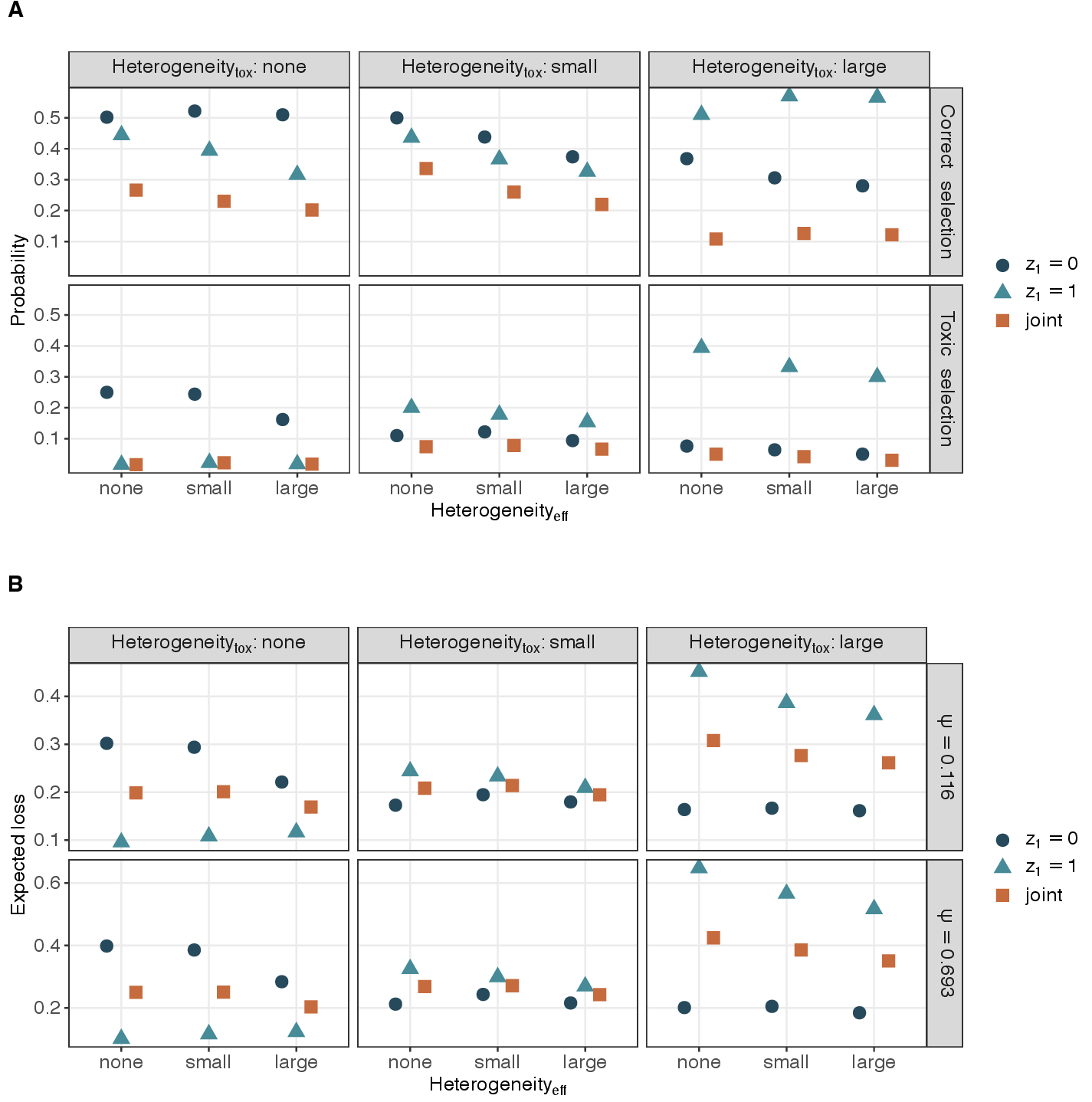}
    \caption{Simulation study selection and expected loss results. The three levels (none, small, large) of heterogeneity in toxicity are denoted by $\text{Heterogeneity}_{tox}$ in the top facet labels of the plots, and the three levels (none, small, large) of heterogeneity in efficacy are denoted by $\text{Heterogeneity}_{eff}$ and are along the x-axis of the plots. In Panel A, the top row of plots corresponds to the probability of correct selection (PCS) and the bottom row of plots corresponds to the probability of toxic selection (PTS) as defined in (\ref{pcs}) and (\ref{pts}), respectively. In Panel B, the top and bottom row of plots correspond to the toxicity constraint violation multipliers for the less ($\psi=0.116)$ and more ($\psi=0.693$) severe violation penalties. The circular and triangular points correspond to the marginal selection probabilities/losses for $z_1=0$ and $z_1=1$ and the square corresponds to the joint selection probability/losses.}
    \label{fig:sim_study_pcs_ptd}
\end{figure}

We let $f(\mathbf{d},z_1)$ and $g(\mathbf{d},z_1)$ be the efficacy and toxicity surfaces which are defined using combinations of the settings provided in Figure \ref{fig:sim_study_eff_tox_settings}. Stratum $z_1=0$ is defined using the first setting for efficacy and toxicity, and then stratum $z_1=1$ is varied to induce the different degrees of heterogeneity in each response surface. The data generating mechanism is $y_f = f(\mathbf{d},z_1) + \epsilon_f$ and $y_g = g(\mathbf{d},z_1) + \epsilon_g$ where $\epsilon_f \sim N(0, \sigma^2_{y_f})$ and $\epsilon_g \sim N(0, \sigma^2_{y_g})$. The specifications for $f(\mathbf{d},z_1)$ and $g(\mathbf{d},z_1)$ under each scenario are included in Table \ref{tbl:sim_study_scenarios} (rows labeled ``Simulation Study'') and $f(\mathbf{d},z_1)$ is visualized in Figure \ref{fig:sim_study_all_scenarios}. The values of $\sigma_{y_f}$ and $\sigma_{y_g}$ are chosen to ensure specific standardized effect sizes ($ses$) at the optimal dose combination, defined as $ses_f=|f_{opt}|/\sigma_{y_f}$ and $ses_g=|g_{opt}|/\sigma_{y_g}$. \cite{willard2023bayesian} considered several values for $ses_f$, which were $0.79/1/3.77$ and represented the $25^{th}/50^{th}/75^{th}$ percentiles of $ses$ from a meta-analysis of dose-responses for a large drug development portfolio at a pharmaceutical company \citep{thomas2014meta}, and showed that a larger $ses$ led to better performance of the algorithm. As we expect these results to generalize to the current setting, we focus specifically on $ses_f=ses_g=1$, which represents the median $ses$ we might expect for a new combination therapy. All computing is performed in R and fully Bayesian inference for the efficacy and toxicity GP models is performed via MCMC in Stan.

To achieve good dose-optimization across the variety of scenarios described above, we need a single model which approximates each scenario well. We describe a method for defining the priors for this model in Section S2 of the Supplementary Materials. This yields the following priors used throughout the simulation study:
\begin{equation}
\begin{aligned}
&\textbf{Efficacy}\ \boldsymbol{f}: \\
&[l_{d_1}/l_{d_2}/l_{z_1}/\nu/\tau] \sim \text{InvGamma}[(11.8, 15.2)/(21, 13.9)/(0.945, 4.47)/(38.8, 16.5)/(575, 545)] \\
&\textbf{Toxicity}\ \boldsymbol{g}: \\
&[l_{d_1}/l_{d_2}/l_{z_1}/\nu/\tau] \sim \text{InvGamma}[(19.5, 16.4)/(18.9, 15.7)/(0.945, 4.47)/(34.6, 16.2)/(575, 545)]
\end{aligned} 
\end{equation}
Three criteria are used to quantify performance of the approach and are estimated using $m=1,...,1000$ Monte Carlo simulations. Probability of correct selection (PCS) measures the probability of selecting the true optimal dose, probability of toxic selection (PTS) measures the probability of selecting a truly toxic dose, and expected loss is used to explicitly characterize the tradeoff between efficacy and toxicity. PCS in stratum $z_k$ is estimated as:
\begin{equation}\label{pcs}
    \text{PCS}_{z_k} \approx \frac{1}{1000}  \sum_{m=1}^{1000} \mathds{1}\{\widehat{\mathbf{d}}_{opt,k}^{(m)} = \mathbf{d}_{opt,k}\}.
\end{equation}
PTS in stratum $z_k$ is estimated as:
\begin{equation}\label{pts}
    \text{PTS}_{z_k} \approx \frac{1}{1000}  \sum_{m=1}^{1000} \mathds{1}\{g(\widehat{\mathbf{d}}_{k,opt}^{(m)}) > 6\}.
\end{equation}
The loss associated with selecting $\widehat{\mathbf{d}}_{opt,k}$ as the optimal dose in stratum $z_k$ is:
\begin{equation}\label{eq:loss_function}
    L_{z_k}^{(m)} = \begin{cases}
        f(\widehat{\mathbf{d}}_{opt,k}^{(m)}) - f(\mathbf{d}_{opt,k}) & \text{if}\ \ g(\widehat{\mathbf{d}}_{opt,k}^{(m)}) \le 6 \\
        \frac{\max_{\mathbf{d}_k \in \mathbb{S}}[f(\mathbf{d}_k) - f(\mathbf{d}_{opt,k})]}{\exp(-\psi(g(\widehat{\mathbf{d}}_{opt,k}^{(m)}) - 6))} & \text{if}\ \ g(\widehat{\mathbf{d}}_{opt,k}^{(m)}) > 6
    \end{cases}
\end{equation}
where $\mathbb{S}=\{\mathbf{d}_k:g(\mathbf{d}_k)\le 6\}$, the set of safe doses, $\psi$ is a toxicity constraint violation multiplier that controls the severity of the penalty for selecting a toxic dose, and $f$ and $g$ are on their original scale. 
This loss function ensures that the loss for selecting a toxic dose exceeds the largest loss among all safe doses. Values of $\psi$ are set by noting that the largest loss among all safe doses is doubled when $\psi=\log(2)/(\Delta - 6)$, where $\Delta >6$. We consider $\Delta\in\{7,12\} \rightarrow \psi\in\{0.693,0.116\}$, where a doubling of the maximum loss among safe doses occurs when toxicity reaches 7 or 12 units, which represents a larger versus smaller penalty for toxic selections, respectively. The expected loss is estimated by averaging $L_{z_k}^{(m)}/\max_{\mathbf{d}_k \in \mathbb{S}}[f(\mathbf{d}_k) - f(\mathbf{d}_{opt,k})]$ over the $m=1,...,1000$ simulations. The divisor ensures that the values for each stratum are on the same scale. We also consider \textit{joint} definitions of these metrics, which are defined as the probability/average loss of simultaneous dose selection across the strata.

Figure \ref{fig:sim_study_pcs_ptd} plots these criteria by toxicity and efficacy scenario. Recall that the response surfaces in stratum $z_1=0$  are fixed throughout, so the degree of response heterogeneity is induced by changing the response surfaces in $z_1=1$. The surfaces in stratum $z_1=1$ always have the same or larger efficacy and the same or larger toxicity as compared to stratum $z_1=0$. For stratum $z_1=0$, we see that PCS, PTS, and expected loss under both values of $\psi$ generally decrease for all efficacy scenarios (comparing efficacy groups along the x-axis across facets) as we move from no to large heterogeneity in toxicity (left to right in the top facet label). For stratum $z_1=1$, we see the opposite trend in PTS and expected loss under both values of $\psi$, which increase as we move from no to large response heterogeneity in toxicity. This results from the number of safe dose combinations in the region of the optimal dose in stratum $z_1=1$ decreasing as we increase toxicity response heterogeneity. For PCS in stratum $z_1=1$, we see a smaller decrease in moving from no to small heterogeneity in toxicity than we did for stratum $z_1=0$, but a large increase when moving from small to large heterogeneity in toxicity. However, this comes at the cost of a large increase in PTS, which is over the 20-30\% range typically deemed acceptable in practice. This corresponds to a large inflation in expected loss under both values of $\psi$ for this scenario. As the locations of the stratum-specific optimal doses become far apart, the joint selection of both optimal doses becomes more challenging, with joint PCS decreasing by half or more when moving from no or small to large heterogeneity in toxicity ($\approx 20-34\%$ vs $\approx 10\%$ joint PCS). Within each toxicity scenario, PCS, PTS, and expected loss for both values of $\psi$ typically decrease as we move from no or small to large heterogeneity in efficacy (groups on x-axis within each facet). This decrease is less pronounced than that resulting from the heterogeneity in toxicity, however, since the locations of the optimal doses do not change. A sensitivity analysis shows that these results are robust to changes in the priors on the kernel hyperparameters but are sensitive to changes in the prior mean function (see Section S4 of the Supplementary Materials).

To conclude, the proposed method achieves good performance given the complexity of the dose optimization -- 25 unique doses and a single optimal dose per stratum. A sensitivity analysis shows that the proposed method also performs better at scaling to different grid sizes as compared to other acquisition function-based approaches (see Section S6 of the Supplementary Materials). Response heterogeneity that leads to a difference in the locations of the optimal doses across strata has a greater impact on PCS, PTS, and expected loss than response heterogeneity which only changes the magnitude of the responses. For this reason, decreased performance is observed when the degree of heterogeneity in toxicity is large. However, when compared to several linear model-based methods using the same set of scenarios, the proposed method is shown to offer better performance on average across all types of response heterogeneity (see Section S5 of the Supplementary Materials). 

\section{Dose-Finding Design for Obstructive Sleep Apnea Therapy}

Obstructive sleep apnea (OSA) is a common sleep disorder which affects over 930 million adults worldwide \citep{schweitzer2023combination} and for which no pharmacotherapy has been approved. One continuous measure used to quantify OSA severity is the apnea-hypopnea index with 4\% oxygen desaturation (AHI$_4$), which is measured in the number of apnea/hypopnea events per hour. Mild to moderate OSA is defined as 10-30 events per hour, and severe OSA is defined as 30 or more events per hour \citep{schweitzer2023combination}. Based on a recently improved understanding of OSA's pathophysiology, a combination of an antimuscarinic agent (aroxybutynin) and a norepinephrine reuptake inhibitor (atomoxetine) has been proposed to serve as a potential pharmacotherapy \citep{schweitzer2023combination}. This study investigated the combination of aroxybutynin (0/2.5/5 mg) and atomoxetine (75 mg) on the reduction in AHI$_4$ from baseline and concluded that the proposed drug combinations display potential as therapy for OSA. While results were not analyzed separately by OSA severity subtype, Figure 2 of \cite{schweitzer2023combination} suggests there is potential response heterogeneity with respect to these subtypes. For the mild to moderate subtype, the combinations (aroxybutynin/atomoxetine) 2.5mg/75mg and 5mg/75mg seem to yield a similar reduction in AHI$_4$, whereas for the severe subtype, the reduction under 5mg/75mg may be greater. However, it was observed that a larger number of AEs resulted from the 5mg/75mg combination, though none were serious. To achieve maximal reduction in AHI$_4$ in the severe subtype, we may be willing to tolerate a larger number of AEs than for the mild to moderate subtype. In this case, the OBD combination would be 2.5mg/75mg for the mild to moderate subtype, but would be the 5mg/75mg combination for the severe subtype. Considering the above, we propose a personalized dose-finding design which tailors the OBD combination to OSA subtype and which permits different toxicity thresholds $g^{\dagger}_k$ for the severe versus mild to moderate subtypes.

To quantify the overall impact of AEs at each dose combination, we follow the approach in \cite{le2020adverse}, who define a non-negative AE burden score that considers both the frequency and severity of the adverse events. \cite{schweitzer2023combination} report the frequency of the most commonly occurring AEs; unfortunately, the grade (severity) information was not provided. However, since no severe adverse events were reported, we define the score by restricting our focus to grades 1-3 only, which we assume to occur 10\%/45\%/45\% of the time independent of the type of AE, reflecting the belief that patients are more likely to remember and report higher graded events. Let $Y_{icg}$ be the indicator that the $i^{\text{th}}$ patient experiences an AE of type $c$ which is of grade $g$, where $g=1,2,3$ and where $c=1,...,8$ corresponding to one of the following types: 1) dry mouth, 2) insomnia, 3) urinary hesitation/flow decrease, 4) constipation, 5) nausea, 6) decreased appetite, 7) feeling jittery, and 8) somnolence. Then patient $i$'s AE burden is $B_i = \sum_c \sum_g w_{cg}Y_{icg}$. Specification of the weights $w_{cg}$ is subjective and should be elicited from subject matter experts. We define $w_{cg}=w_cw_g$ and set grade weights $w_g=g$. Higher type weights $w_c$ may be given to more burdensome types of AEs if desired. As was noted in \cite{schweitzer2023combination}, the AEs which led to study discontinuation included insomnia, nausea, and dry mouth. We assign these AEs weights which are 5 times higher than the others and thus $\mathbf{w}_{c}=(5,5,1,1,5,1,1,1)$. Specifying different weights would change the dose-toxicity surface described below. However, we would expect our method to be robust to changes in the dose-toxicity surface given its good performance under a variety of toxicity scenarios investigated in the simulation study above. Interestingly, it is suggested in \cite{schweitzer2023combination} that the 2.5mg/75mg combination decreased the frequency of AEs overall, as well as the frequency of more burdensome AEs as compared to the 0mg/75mg combination. This non-monotoniticy in the dose-toxicity relationship is captured in the simulations that follow.

\begin{figure}
    \centering
    \includegraphics[width = \linewidth]{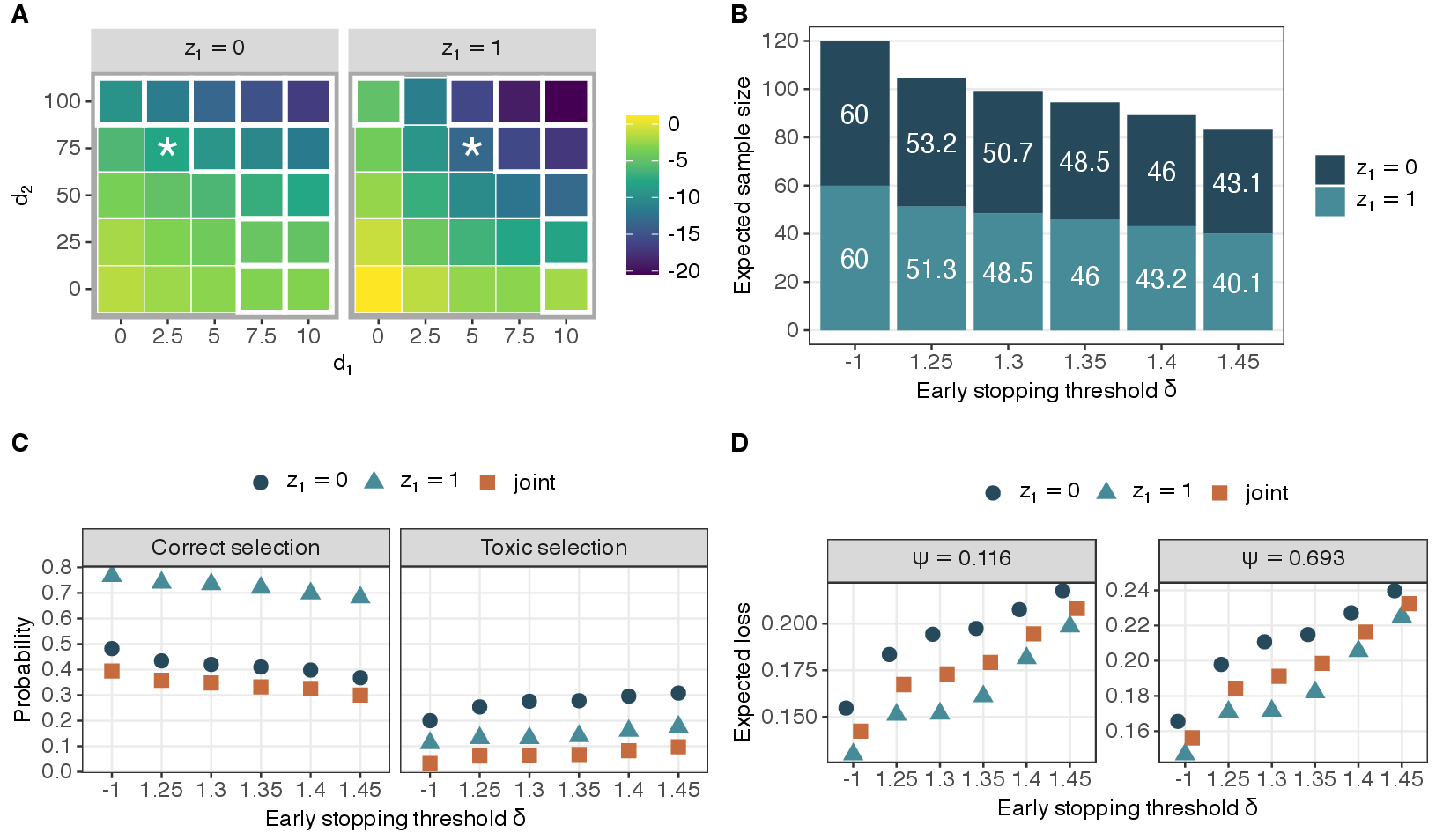}
    \caption{A) Efficacy surface, white stars denote $\mathbf{d}_{opt,k}$, white borders around cells denote toxic doses which are doses where $g(\mathbf{d},z_1=0) > 1.5$ and $g(\mathbf{d},z_1=1) > 2$, B) expected sample size, C) selection probabilities including probability of correct selection (PCS) and probability of toxic selection (PTS) as defined in (\ref{pcs}) and (\ref{pts}), respectively, and D) expected loss using the loss function defined in (\ref{eq:loss_function}) for two values of the toxicity constraint violation multiplier $\psi$. The circular and triangular points correspond to the marginal selection probabilities/loss for $z_1=0$ and $z_1=1$ and the square corresponds to the joint selection probability/loss. For OSA subtype, $z_1=0$ denotes mild/moderate and $z_1=1$ denotes severe. }
    \label{fig:rwa}
\end{figure}

We define $z_1$ as OSA severity, where $z_1=0$ corresponds to the mild/moderate subtype and $z_1=1$ corresponds to the severe subtype. We combine two dosing agents, aroxybutynin and atomoxetine, where aroxybutynin is denoted by $d_1$ with dose levels $\{0, 2.5, 5, 7.5\}$ in mg and where atomoxetine is denoted by $d_2$ with dose levels $\{0, 25, 50, 75, 100\}$ in mg. We consider the 25 unique dose combinations that result from the Cartesian product of $d_1$ and $d_2$. We define  $f(\mathbf{d},z_1)$ as the reduction in AHI$_4$ from baseline and $g(\mathbf{d},z_1) = \log(B + 0.5)$ as the log-transformed AE burden score. Using efficacy and adverse event results from \cite{schweitzer2023combination}, we fit flexible linear models on standardized doses to determine the parameters used in the data generating mechanisms for $f(\mathbf{d},z_1)$ and $g(\mathbf{d},z_1)$, whose functional forms are included in Table \ref{tbl:sim_study_scenarios} (rows labeled ``OSA'') and which are plotted in Panel A of Figure \ref{fig:rwa} on the original dosage scale. The $f(\mathbf{d},z_1)$ differ across the strata, where the parameter vector for $z_1=0$ is $\boldsymbol{\phi}_{z_1=0}=(-1.38, -4.08, -0.48, -4.23, 2.45, -7.51, -1.56)$ and the parameter vector for $z_1=1$ is $\boldsymbol{\phi}_{z_1=1}=(1.05, -11.28, -8.32, -17.02, 8.17, 2.34, 4.61)$. The $g(\mathbf{d},z_1)$ is the same across strata and has parameter vector $\boldsymbol{\psi}=(-0.59, 1.83, 2.26, -4.05, 1.79, 0.47, 2.91)$. As mentioned previously, we permit the tolerable toxicity thresholds to vary by strata. For the severe subtype, we set $g^{\dagger}_{1}=2$, which equates to accepting an average AE burden score of about 7. This may include, for example, a single low-grade but high burden AE, or a couple moderate-grade but low burden AEs on average. For the mild/moderate subtype, we set $g^{\dagger}_{0}=1.5$, which equates to accepting an average AE burden score of about 4. This may include, for example, a couple low-to-moderate-grade AEs of low burden, but would exclude higher burden AEs on average. 

Below we compare 6 designs using a maximum sample size of 120 and where the number of patients assigned to each dose combination is fixed at $r_{k,t}=2$ for all $t$ and the rate of expansion of the dose combination region is set at $\rho_k=0.5$. The 6 designs differ in when they permit the trial to stop early for efficacy, where we use the posterior entropy stopping rule and thresholds $\delta_k$ previously described. Recall $H_k<\delta_k$ must be satisfied at least $(J+1)$ times for early stopping for efficacy to occur, which, for the dual-agent combination therapy being investigated, equates to 3 times. For each design, we use the same value of $\delta_k$ for each stratum, and so $\delta_k=\delta$. Thus the 6 designs are defined using $\delta = \{-1,1.25,1.3,1.35,1.4,1.45\}$. No early stopping is permitted when $\delta =-1$, and increasingly earlier stopping is permitted as we move from $\delta=1.25$ to $\delta=1.45$. Furthermore, the algorithm permits stratum-specific early stopping, where the remaining budget is allocated to the other stratum where dose-finding may continue in the event of one of the strata stopping early. Finally, as there are no lethal or serious adverse events expected, we do not permit early stopping for toxicity. We set our priors in the same manner as described in Section S2 of the Supplementary Materials but for the single data generating mechanism defined above, yielding the following values for $\boldsymbol{\beta}$ (Figure \ref{fig:prior_mean}) and the GP kernel hyperparameters $\boldsymbol{\theta}$:
\begin{equation}
\begin{aligned}
&\textbf{Efficacy}\ \boldsymbol{f}: \\
&[l_{d_1}/l_{d_2}/l_{z_1}/\nu/\tau] \sim \text{InvGamma}[(28.0, 25.3)/(30.7,23.5)/(0.945,4.47)/(53.8,44.1)/(461,382)] \\
&\textbf{Toxicity}\ \boldsymbol{g}: \\
&[l_{d_1}/l_{d_2}/l_{z_1}/\nu/\tau] \sim \text{InvGamma}[(26.4,18.8)/(17.7,17.8)/(0.945,4.47)/(49.5,63.3)/(449,325)]
\end{aligned} 
\end{equation}
All other modeling and inferential details follow those previously described in the simulation study section. Design performance is assessed via the previously defined criteria which are estimated using 1,000 Monte Carlo replicates after setting $g^{\dagger}_{1}=2$ and $g^{\dagger}_{0}=1.5$.  
Under this scenario, we find that earlier stopping leads to decreased expected sample size, as expected (Panel B of Figure \ref{fig:rwa}). Under no early stopping, we find PCS to be 77\% and 48\% for strata $z_1=1$ and $z_1=0$, respectively, and that the joint PCS is 40\% (Panel C of Figure \ref{fig:rwa}). PTS is at most 20\% for all cases under no early stopping. When early stopping for efficacy is permitted, we see a reduction in PCS and an increase in PTS for all cases, with a greater reduction/increase the earlier the algorithm terminates. We see similar trends in expected loss for both values of $\psi$. Performance is better in stratum $z_1=1$, having both a higher PCS and lower PTS as well as lower expected loss. This is explained by the $ses$'s being larger in this stratum. This permits the algorithm to stop earlier in stratum $z_1=1$ and place additional patients in the more challenging stratum $z_1=0$, which results in a greater expected sample size in $z_1=0$ (Panel B of Figure \ref{fig:rwa}). If we were to tolerate at most 30\% PTS, then we would recommend designs $\delta=\{-1,1.25,1.3,1.35\}$ for further consideration by the sponsor, who could then select a final design by weighing the tradeoffs between PCS, PTS, expected loss, and expected sample size.

\section{Discussion}

In this work, we generalized the method proposed in \cite{willard2023bayesian} to the setting of higher-grade continuous toxicities under no monotonicity assumptions. We described a dose-escalation scheme and showed how toxicity information could be incorporated into the search strategy employed by the Bayesian optimization methods. We proposed a novel adaptive stopping rule and demonstrated how to perform inference in a fully Bayesian manner. We conclude that the approach performs well and can efficiently identify stratum-specific optimal doses in the presence of response heterogeneity.

The proposed work is not without limitations. Firstly, all simulations were performed using a single binary covariate under the setting of dual-agent dose combinations only. The use of additional categorical variables or dosing agents is straightforward, though would require larger sample sizes. Extension to continuous covariates remains as future work. While conceptually straightforward under the proposed approach, the consequences of inadvertently extrapolating into regions of covariate combinations never before seen in the trial would need to be considered and assessed. On a related point, the simulations assumed that the patient subgroups defined by the strata were equally prevalent and had equal enrollment rates, thus ensuring extrapolation was not an issue. Dose optimization would likely be impacted if some of the strata are sparse or have slower enrollment rates. \cite{zhang2024adaptive} cautions against stratum-specific dose optimization when the numbers of patients in each subgroup are expected to be very different. A full investigation of the impact of unequal prevalence and enrollment rates across the strata remains for later exploration. Finally, we assumed conditional independence between the efficacy and toxicity surfaces given dose and covariates. In trials with binary efficacy and toxicity responses, \cite{cunanan2014evaluating} showed that estimating correlations can be challenging given the small sample sizes of early phase designs and that designs which assumed independence between efficacy and toxicity can still perform well. Indeed, in a simulation study in Section S2 of the Supplementary Materials, we show that the performance of the proposed method is fairly robust to even quite large degrees of correlation. However, additional performance gains may be possible by relaxing the assumption of conditional independence. A multivariate GP could jointly model the efficacy and toxicity responses and the cEI acquisition function could be modified to incorporate the correlation between the two responses. We leave this as a direction for future work. 

Changes to the GP modeling may help improve performance of the algorithms. Stationary, anisotropic squared exponential kernels were used but could be replaced by \textit{non-stationary} kernels which include dose-covariate interaction terms. These would permit the correlation between two dose combinations to depend on the actual dosage levels and strata in which they lie, rather than solely on their distance from one another in the input space.  Doing so, however, may greatly increase the dimensionality of the hyperparameter space and make estimation more challenging.


\backmatter


\section*{Acknowledgments}
The authors would like to thank Dr. John Kimoff for his valued input regarding the OSA application component of this manuscript, as well as the anonymous associate editor and two reviewers whose comments helped improve the manuscript. JW acknowledges the support of a doctoral training scholarship from the Fonds de recherche du Québec - Nature et technologies (FRQNT), grant number MC\_UU\_00040/3, and notes that this report is independent research supported by the National Institute for Health Research (NIHR300576). The views expressed in this publication are those of the authors and not necessarily those of the NHS, the National Institute for Health Research, or the Department of Health and Social Care (DHSC). SG acknowledges the support by a Discovery Grant from the Natural Sciences and Engineering Research Council of Canada (NSERC), a Chercheurs-boursiers (Junior 1) award from the Fonds de recherche du Québec - Santé (FRQS), and support from the Canadian Statistical Sciences Institute. EEMM acknowledges support from a Discovery Grant from NSERC. EEMM is a Canada Research Chair (Tier 1) in Statistical Methods for Precision Medicine and acknowledges the support of a chercheur de mérite career award from the FRQS. This research was enabled in part by support provided by Calcul Québec, the Digital Research Alliance of Canada, and the University of Cambridge Research Computing Service.\vspace{-10pt}

\section*{Supplementary Materials}
Web Appendices referenced in Sections 2-5 and data and code to run the simulation studies and produce the graphics included in the manuscript are available with this paper at the Biometrics website on Oxford Academic.
\vspace{-10pt}

\section*{Data Availability}
No new data were created or analyzed in this study.\vspace{-10pt}






\bibliographystyle{biom}
\bibliography{references}
\vspace*{-10pt}








\label{lastpage}

\end{document}